\documentstyle[12pt]{article} 
\parskip=\medskipamount
\def\be{\begin{equation}}
\def\ee{\end{equation}}

\newcounter{fig}

\begin{document}

\begin{titlepage}

\begin{center}

{\Huge \bf  Biased diffusion in a one-dimensional 
adsorbed monolayer 
}

\vspace{0.8in}

{\Large \bf O.B\'enichou$^{1}$, A.M.Cazabat$^{2}$, A.Lemarchand$^{1}$, \\
M.Moreau$^{1}$
 and G.Oshanin$^{1}$
}

\vspace{0.5in}

{\large \sl $^{1}$ Laboratoire de Physique Th\'eorique des Liquides (CNRS - UMR 7600), 
Universit\'e Pierre et Marie Curie,
4 place Jussieu, 75252 Paris Cedex 05, France\\
}
\vspace{0.3in}
{\large \sl $^{2}$ Laboratoire de Physique de la Mati\`ere Condens\'ee,
Coll\`ege de France, 11 place M.Berthelot, 75252 Paris Cedex 05, France
} 

\vspace{0.8in}

\begin{abstract}

We study dynamics of a probe particle, which performs biased diffusive motion in 
a one-dimensional adsorbed  monolayer of mobile
hard-core particles undergoing continuous exchanges 
with a vapor phase. In terms of  a mean-field-type
approach, based on the decoupling of the third-order correlation functions 
into a product of the
pairwise correlations,  
we determine analytically the
density profiles of the monolayer particles, 
as seen from the stationary moving probe,  and
calculate
the terminal velocity $V_{pr}$, mobility $\mu_{pr}$ and the
 self-diffusion coefficient $D_{pr}$ of the probe.
Our analytical results are confirmed by Monte Carlo simulations.

\end{abstract}

\vspace{0.8in}

\end{center}

\end{titlepage}

\pagebreak

\section{Introduction}

When a solid surface is brought in contact with an ambient gas phase,
 the interactions
between the molecules of both systems often result in formation
of a layer of gas particles covering the solid surface. 
Following a seminal work of Langmuir (see, e.g., in  Ref.\cite{surf}),  
who has invented a somewhat simplified description
in which 
the interactions between the adsorbed molecules 
were regarded as a mere hard-core,  
thermodynamic properties of
such adsorbed layers, as well as different forms of possible 
phase transformations 
have been extensively studied 
and a number of
important developments have been made. 
In particular, 
subsequent studies included more realistic forms of  
intermolecular interactions or allowed for the possibility of
multilayer formation.  In consequence, more sophisticated
 forms of adsorption isotherms 
have been evaluated  which explain quite
well available experimental data 
(see, e.g.  Refs.\cite{surf,hill,fowler}).  

Considerable effort has been also invested in
understanding of molecular diffusion in adsorbed layers,  
which has a strong impact on their properties  
\cite{gomer,kreuzer}.
Here, some approximate results have been obtained for
both dynamics of an isolated adatom
on a corrugated surface and collective
diffusion, describing spreading of the macroscopic 
density fluctuations in
 interacting adsorbates being in contact with 
the vapor phase
\cite{gomer,kreuzer,kotur,gortel,ala,kehr}. 
On the other
hand, 
available studies of the
tracer diffusion  
in adsorbed layers, 
which is observed experimentally in  
STM or field ion measurements and 
provides
a useful information about such properties of monolayers 
as, e.g., their intrinsic viscosity,
pertain to  stricktly two-dimensional models excluding the possibility 
of particles adsoption or desorption (see, e.g. Refs.\cite{ala,kehr,deconinck}
and references therein). 
Analysis of the tracer diffusion and mobility of impure molecules 
in adsorbed monolayers undergoing exchanges with the vapor phase
 seems to be lacking at present.

In this paper we study the time evolution of
 a model system consisting of
a solid substrate covered by a monolayer of mobile
hard-core particles undergoing continuous exchanges 
with a vapor,  
and a single
impure, probe particle,  which is subject to
 a constant external force $E$
and hence performs
a biased random walk constrained by hard-core
 interactions with the
 monolayer particles.
Here, we concentrate on the one-dimensional
 case and model, in a usual
 fashion, the solid substrate as a regular 
one-dimensional lattice of
 adsorbing sites, which can support, at most,
 a single occupancy; results 
for the two-dimensional monolayer will be
 published elsewhere. 
In terms of a mean-field-type
approach of Ref.\cite{burlatsky}, which pressumes
 certain decoupling of 
the third-order correlation functions, 
we define the
density profiles of the monolayer particles, 
as seen from the stationary moving probe, and
determine
analytically
the probe terminal velocity $V_{pr}(E)$. In the most general case 
$V_{pr}(E)$ is obtained implicitly, as a solution of a non-linear 
equation relating its value to the system parameters. This equation 
simplifies  
in the limit of small $E$ and here 
the probe velocity can be found explicitly; 
we show that it
obeys $V_{pr}(E) \approx E/\zeta$, where 
the friction coefficient $\zeta$, which is inverse of the probe mobility
$\mu_{pr}$, is expressed through 
the microscopic parameters characterizing the system under study. 
This result establishes the frictional drag force exerted on the probe
by the monolayer particles in the low-$E$ limit and thus can be thought off as
the analog of the Stokes' law for the one-dimensional monolayer in contact with vapor.
Lastly, 
we determine  the
 self-diffusion coefficient $D_{pr}$ of the probe, which is computed here
by
assuming heuristically 
the validity of the Einstein relation 
between the self-diffusion
coefficient and the mobility.
Our analytical results for the probe terminal velocity and
 the self-diffusion coefficient, as
well as  for the stationary density profiles around it 
are confirmed by Monte Carlo simulations 
of the 
corresponding master equation by the method of Gillespie \cite{gillespie}.

The paper is structured as follows: In Section
 2 we formulate
the model and introduce
basic notations.  In Section 3 we write down the dynamical
equations which govern the time evolution of the monolayer
particles and of the probe. Sections 4 and 5 are  devoted to the
analytical solutions of these
evolution equations in the continuous-space limit
 and on the discrete lattice, respectively. In these sections we also present a 
comparison of the analytical results and
Monte Carlo simulations data.
Finally, we conclude in Section 6 with a brief summary
and discussion of our results.

\section{The model}

The model consists of a one-dimensional regular,  
infinite in both directions
lattice of spacing
$\sigma$,  
the sites $X$ of which are 
connected to a reservoir (a vapor phase) containing an infinite number 
of identical, electrically neutral 
gas particles (Fig.1). The particles from the reservoir 
may adsorb onto the lattice sites, desorb from them
or move along the lattice by performing symmetric 
random walk between the
neighboring sites;  
adsorption onto the lattice and hops
between the lattice sites are constrained by hard-core exclusion -
each lattice site can be 
either singly occupied  or vacant.  The state of each lattice site $X$
at time $t$ is described by time-dependent occupation variable
$\eta(X)$, which can take two values
\begin{equation}
\eta(X) = \left\{\begin{array}{ll}
1,     \mbox{ if the site $X$ is occupied by a gas particle} \nonumber\\
0,     \mbox{ otherwise}
\end{array}
\right.
\end{equation}
As one may readily notice, the just-described system corresponds to
a one-dimensional
version of the Langmuir adsorption/desorption model,
 with the only difference being that
the  lateral diffusion of the adsorbed molecules is allowed.
Note, however, that the possibility of 
particles lateral diffusion makes the model more
complicated, compared the Langmuir's one, since now the evolution of particle local
densities at different sites is coupled due to diffusion.

Further on, we place at the site $X = 0$ at time $t = 0$
an extra particle, 
which will be referred to in what follows as the "probe",
 since it allows us to probe the
resistance offered by the adsorbed layer to an external 
perturbance. Position of this particle at time $t$ for a given
realization of the process will be denoted as
$X_{pr}(t)$.
We stipulate that this particle is different from 
the monolayer  particles in that it can not desorb back into the vapor
phase,
i.e. is constrained to move along the lattice only.  
Second,  we suppose
that this only particle is charged (for simplicity, we set the charge equal to unity
in what follows) and 
is subject to a constant external
electric field $E$, which favors its motion in a preferential direction.
The questions which we address here are, first,  the dependence
of the probe particle terminal velocity $V_{pr}(E) = lim_{t \to \infty} V_{pr}(t)$
on the magnitude of the driving force $E$ and other system parameters;  second, the 
form of the density profiles as seen from the stationary moving probe, and lastly, the
self-diffusion coefficient of the probe in absence of the driving force.

Now, we define particle dynamics and system parameters more precisely:

(a) $\em Diffusion \; and \; desorption$. 
Each of the adsorbed at time moment $t$
 gas particles waits a random, exponentially
distributed time with mean  $\tau^{*}$, and then selects between
 either of three possibilities; it may choose to leave 
the lattice with a probability $g$, or attempt to hop,
with the probability \( (1-g)/2 \), to one of the two 
neighboring sites.  If the
desorption event is chosen - the particle leaves the lattice
instantaneously.
On the other hand, in case when 
the particle attempts to hop to one of the neighboring sites,
the jump can actually occur only if the target site 
is empty; otherwise, the particle remains at its position. 

(b) $\em Adsorption. $
 Particles of the reservoir wait a random,
exponentially distributed time with mean\footnote{We choose
for simplicity the same mean waiting time for the particles
of the lattice and those in  the reservoir; our results
can be simply generalized for arbitrary $\tau_{ad}$ by replacement 
$ f \to \tau^{*} f/\tau_{ad}$.} $\tau_{ad} = \tau^{*} $  and then
attempt to adsorb onto the lattice with a
probability f. As previously, 
an adsorption event takes place only 
if the chosen site is 
empty. 

Note,
that the total number of particles on the lattice is not conserved
in  such a dynamics. Mean density $\rho =  <\eta(X)>$, however, 
approaches as $t \to \infty$ a constant value 
\begin{equation}
\rho_s =  \frac{f}{f + g}, 
\label{langmuir}
\end{equation}
which  relation is often called in the literature
as the  Langmuir adsorption isotherm \cite{surf}. Here we suppose that
 parameters $f$ and $g$ are independent of each other 
and may take arbitrary values from the interval $[0;1]$. 
In reality, their values are
prescribed 
by the pressure of the gas phase, 
precise form of the solid-gas interaction potential
and the temperature. As well, 
the latter determine the characteristic 
time $\tau^{*}$,
which is also considered here 
as an independent given parameter.

(c) $\em Dynamics \; of  \; the \;  probe.$ 
The probe particle waits
an exponentially distributed time with mean $\tau$, 
(which can be, in general case,
different of  $\tau^{*}$) and then
selects, at random,  a jump direction: It chooses a right-hand or
left-hand adjacent site with probabilities $p$ and  $q=1-p$, respectively. 
Similarly to the gas particles, the jump is only then fulfilled when 
 the selected site is vacant at this moment of time.  

In a usual fashion, 
the 
jump probabilities are related to the external electric field $E$, (oriented in the positive direction, such that \(E\geq 0\)),
and the reciprocal temperature $\beta$
by 
\begin{equation}
p/q = \exp(\beta \sigma E),
\end{equation}
where we have set, for simplicity, the probe charge equal to 
unity. This relation  together with the condition $p + q = 1$ defines
 the values of $p$ and $q$. 

\section{Evolution equations}

Let $P(X_{pr},\eta;t)$ denote the probability of finding  at time $t$ the probe at the site $X_{pr}$
and all other adsorbed particles in the configuration $\eta = \{\eta(X)\}$. 
Further on, let  $\eta^{z,z+1}$  be 
 the configuration obtained from $\eta$ by exchanging
the occupation variables of sites $X = z$ and $X = z + 1$, i.e.,  
$\eta(z) \leftrightarrow \eta(z+1)$,
and $\eta^{z}$ - the configuration obtained from $\eta$ by replacement
$\eta(X=z) \to 1-\eta(X=z)$.
Then, summing up all events which may change or result in a configuration $(X_{pr},\eta)$,
we  find that the time
evolution of $P(X_{pr},\eta;t)$ is guided by
 the following master equation:
\begin{eqnarray}
\dot{P}(X_{pr},\eta;t)&=&\frac{1-g}{2\tau^*}\sum_{z\neq X_{pr}-\sigma,X_{pr}}
\Big\{P(X_{pr},\eta^{z,z+1};t)-P(X_{pr},\eta;t)\Big\}\nonumber\\
&+&\frac{p}{\tau}\Big\{(1-\eta(X_{pr}))P(X_{pr}-\sigma,\eta;t)-(1-\eta(X_{pr}+\sigma))P(X_{pr},\eta;t)\Big\}\nonumber\\
&+&\frac{q}{\tau}\Big\{(1-\eta(X_{pr}))P(X_{pr}+\sigma,\eta;t)-(1-\eta(X_{pr}-\sigma))P(X_{pr},\eta;t)\Big\}\nonumber\\
&+&\frac{g}{\tau^*}\sum_{z\neq X_{pr}}\Big\{(1-\eta(z))P(X_{pr},\eta^z;t)-\eta(z)
P(X_{pr},\eta;t)\Big\}\nonumber\\
&+&\frac{f}{\tau^*}\sum_{z\neq X_{pr}}\Big\{\eta(z)P(X_{pr},\eta^z;t)-(1-\eta(z))
P(X_{pr},\eta;t)\Big\}
\label{eqmaitresse}
\end{eqnarray}
where the dot denotes the time derivative, 
 the terms in the first three lines describe respectively the diffusion of
the adsorbed gas particles
and of the probe, which proceed due to the
Kawasaki-type particle-vacancy exchanges, while 
the last two lines account respectively for the Glauber-type
desorption and adsorption processes.

Equation (\ref{eqmaitresse}) allows to compute
 the instantaneous velocity of the probe molecule.
Multiplying both sides of Eq.(\ref{eqmaitresse}) by $X_{pr}$, and summing over all
possible states $(X_{pr},\eta)$ of the system,
we find that :
\begin{equation}
V_{pr}(t) = \frac{d \overline{X_{pr}(t)}}{d t} = \frac{\sigma}{\tau}\Big\{p(1-k(\sigma;t))-
q(1-k(-\sigma;t))\Big\}
\label{vitesse}
\end{equation}
where $\overline{X_{pr}(t)}$ denotes the mean displacement of the probe at time $t$ and
\begin{equation}
k(\lambda;t)=\sum_{(X_{pr},\eta)}\eta(X_{pr}+\lambda)P(X_{pr},\eta;t)
\label{fonccorreldef}
\end{equation}
defines the probability of having at time t
an adsorbed gas particle at distance $\lambda$ from the probe, or in other 
words, can be interpreted as the monolayer 
density profile as seen from the
moving probe.

Hence, in order to compute $V_{pr}(t)$ we have to determine 
the evolution of $k(\lambda;t)$.
From Eq.(\ref{eqmaitresse}) it follows that we have to consider separately
the evolution of $k(\lambda;t)$ for $|\lambda| > \sigma$ and $|\lambda|= \sigma$,
since the monolayer particles dynamics is different in these two domains.
We start first with the case 
$|\lambda| > \sigma$, in which domain the evolution of $k(\lambda;t)$ is not 
directly affected by the presence of
the probe.  In this case we find from Eq.(\ref{eqmaitresse}) the following equation
\begin{eqnarray}
\dot{k}(\lambda;t)&=&\frac{1-g}{2\tau^*}\Big\{k(\lambda+\sigma;t)
+k(\lambda-\sigma;t)-2k(\lambda;t)\Big\}-
\frac{f+g}{\tau^*}k(\lambda;t)+\frac{f}{\tau^*}+\nonumber\\
&+&\frac{p}{\tau}
\sum_{(X_{pr},\eta)} (1-\eta(X_{pr}+\sigma))P(X_{pr},\eta;t)\Big\{\eta(X_{pr}+\lambda+\sigma)-\eta(X_{pr}+\lambda)\Big\}-\nonumber\\
&-&\frac{q}{\tau}\sum_{(X_{pr},\eta)} (1-\eta(X_{pr}-\sigma))P(X_{pr},\eta;t)\Big\{\eta(X_{pr}+\lambda)-\eta(X_{pr}+\lambda-\sigma)\Big\},
\label{e}
\end{eqnarray}
where the terms in the first  line represent the contributions due to
 the particles diffusion, desorption and
adsorption, while the terms in the last two lines are associated with the translation
of the configuration $\eta \to \eta^{z,z \pm 1}$ due to the backward and forward
hops of the probe particle. These terms are non-linear with respect 
to the occupation variables and thus couple the
evolution of the "pairwise" correlation function $k(\lambda;t)$ to 
the evolution of the third-order correlations.
Thus, to define the behavior of $k(\lambda;t)$ one faces the problem of solving
an infinite hierarchy of coupled differential equations
for the higher-order correlation functions.

Here we will resort to an approximate, mean-field-type decoupling scheme, 
which has been first applied
in Ref.\cite{burlatsky} to describe tracer diffusion in a one-dimensional
 hard-core lattice gas with conserved number of
particles,
i.e. a gas for which the exchanges with the reservoir are forbidden
 and both $f$ and $g$ are equal to zero.  
It has been shown in Ref.\cite{burlatsky} 
that results of such an approach 
are in a very good  agreement with the
 Monte Carlo simulations data. Moreover,  
rigorous probabilistic analysis \cite{olla} of this model
produced essentially the same results as Ref.\cite{burlatsky}, thus proving that 
such a decoupling scheme renders exact description of the model.
We note also parenthetically that
 a very good agreement between the numerical data 
and analytical predictions, based on the same 
decoupling scheme, have been observed for a 
slightly different model of hard-core gas spreading 
on a one-dimensional lattice from a reservoir connected
 to one of the lattice sites \cite{spreading}. 
 We thus adopt here this mean-field-type 
approach, which provides quite a good description of related dynamical models,
 and will verify in what follows
our analytical predictions against the results of Monte Carlo simulations.

The decoupling scheme of Ref.\cite{burlatsky} 
is based on the assumption that  the average with the weight 
$P(X_{pr},\eta;t)$ of the product of several occupation variables 
of different sites 
factorizes into the product of their average values with the weight $P(X_{pr},\eta;t)$. Namely, it assumes 
that  the average product of the occupation variables factorizes as 
\begin{eqnarray}
&&\sum_{(X_{pr},\eta)} \eta(X_{pr}+\lambda)(1-\eta(X_{pr}\pm\sigma))P(X_{pr},\eta;t)=\nonumber\\
&=&\Big\{\sum_{(X_{pr},\eta)}\eta(X_{pr}+\lambda)P(X_{pr},\eta;t)\Big\}
\times\Big\{\sum_{(X_{pr},\eta)}(1-\eta(X_{pr}\pm\sigma))P(X_{pr},\eta;t)\Big\}=\nonumber\\
&=&k(\lambda;t)(1-k(\pm\sigma;t))
\label{decouplage}
\end{eqnarray}
Then, taking advantage of Eq.(\ref{decouplage}), we can rewrite Eq.(\ref{e}), which holds for $|\lambda| > \sigma$,  in
the following form:
\begin{eqnarray}
\dot{k}(\lambda;t)&=&\frac{1-g}{2\tau^*}\{k(\lambda+\sigma;t)
+k(\lambda-\sigma;t)-2k(\lambda;t)\} -\frac{f+g}{\tau^*}k(\lambda;t)+\frac{f}{\tau^*}+\nonumber\\
&+&\frac{p}{\tau}\{1-k(\sigma;t)\}\{k(\lambda+\sigma;t)-k(\lambda;t)\}
-\frac{q}{\tau}\{1-k(-\sigma;t)\}\{k(\lambda;t)-k(\lambda-\sigma;t)\}
\label{eqpourklambda}
\end{eqnarray} 
Note, however, that despite the fact that the 
decoupling in Eq.(\ref{decouplage}) allows us to close
the hierarchy in Eq.(\ref{e}) at the level of pairwise correlations, 
the resulting equations still pose some technical problems 
for solving them; namely,  they are  non-linear with respect
to $k(\lambda;t)$ differential 
equations. The method of solution will be discussed
in the next two sections.

Now, similar analysis can be carried out to derive 
the dynamical equations, which govern the evolution
of $k(\lambda;t)$ at points $|\lambda| = \sigma$.  We find, respectively,
\begin{eqnarray}
\dot{k}(\sigma;t)&=&\frac{1-g}{2\tau^{*}}\Big\{k(2\sigma;t)-k(\sigma;t)\Big\}-
\frac{f+g}{\tau^{*}}k(\sigma;t)+\frac{f}{\tau^{*}}+\nonumber\\
&+&\frac{1}{\tau}\Big\{-qk(\sigma;t)(1-k(-\sigma;t))
+p(1-k(\sigma;t))k(2\sigma;t)\Big\}
\label{eqpourksigma}
\end{eqnarray}
and
\begin{eqnarray}
\dot{k}(-\sigma;t)&=&\frac{1-g}{2\tau^{*}}\Big\{k(-2\sigma;t)-k(-\sigma;t)\Big\}
-\frac{f+g}{\tau^{*}}k(-\sigma;t)+\frac{f}{\tau^{*}}\nonumber\\
&+&\frac{1}{\tau}\Big\{-pk(-\sigma;t)(1-k(\sigma;t))
+q(1-k(-\sigma;t))k(-2\sigma;t)\Big\}
\label{eqpourk-sigma}
\end{eqnarray}
Equations (\ref{vitesse}),(\ref{eqpourklambda}) and (\ref{eqpourksigma}),(\ref{eqpourk-sigma}) 
constitute a closed system of
non-linear equations, which suffice the computation of the probe velocity and other characteristic properties.

\section{\bf Stationary solution of the  evolution equations in the conti-nuous-space limit.}

Consider first the solution to Eqs.(\ref{vitesse}),(\ref{eqpourklambda}) 
and (\ref{eqpourksigma}),(\ref{eqpourk-sigma}) in the continuous-space limit. Expanding 
$k(\lambda\pm\sigma;t)$ in Taylor series  up to the second order in powers of $\sigma$
(diffusion limit),
we have that $k(\lambda;t)$ with $|\lambda| > 0$ obeys:
\begin{equation}
\dot{k}(\lambda;t)=D_0 \frac{\partial^2 k(\lambda;t) }{\partial \lambda^2} 
+ V_{pr}(t) \frac{\partial k(\lambda;t) }{\partial \lambda}
-\frac{f+g}{\tau^*}k(\lambda;t)+\frac{f}{\tau^*},
\label{cslambda}
\end{equation}
where $D_0$ denotes the "bare" diffusion coefficient of the adsorbed
gas particles, $D_0 = (1 - g) \sigma^2/2 \tau^{*}$,
and the velocity $V_{pr}(t)$ is now given by
\begin{equation}
V_{pr}(t)=\frac{p \sigma}{\tau} \Big\{1-k(\lambda=+0;t)\Big\}-\frac{q \sigma}{\tau} 
\Big\{1-k(\lambda=-0;t)\Big\}
\label{newvel}
\end{equation}
Similarly, we find that Eqs.(\ref{eqpourklambda}) and (\ref{eqpourksigma}) take the form
\begin{equation}
\sigma \dot{k}(\lambda=+0;t)=D_0 \left. \frac{\partial k(\lambda;t) }{\partial \lambda} \right|_{\lambda=+0}
+ \Big\{V_{pr}(t) - (f + g) \frac{\sigma}{\tau^{*}}\Big\}k(\lambda=+0;t)
+\frac{\sigma f}{\tau^*},
\label{cslambda0}
\end{equation}
and
\begin{equation}
\sigma \dot{k}(\lambda=-0;t)=-D_0 \left. \frac{\partial k(\lambda;t) }{\partial \lambda} \right|_{\lambda=-0}
- \Big\{V_{pr}(t) + (f + g) \frac{\sigma}{\tau^{*}}\Big\}k(\lambda=-0;t)
+\frac{\sigma f}{\tau^*}
\label{cslambda-0}
\end{equation}

We turn next to the limit $t \to \infty$. Assuming first that the probe terminal velocity 
 approaches some constant value 
$V_{pr}(E)$ as $t \to \infty$, i.e. $V_{pr}(E)
 = lim_{t \to \infty} V_{pr}(t)$,  we  find 
the stationary solution $k(\lambda) = lim_{t \to \infty} k(\lambda;t)$
of Eqs.(\ref{cslambda}),(\ref{cslambda0}) and (\ref{cslambda-0}). This reads:
\begin{equation}
k(\lambda)=\rho_s \Big[1+A_{\pm} \exp\Big(-|\lambda|/\lambda_{\pm}\Big)\Big]
\label{profiles}
\end{equation}
In Eq.(\ref{profiles}) 
$\rho_s$ is determined by Eq.(\ref{langmuir}), 
the sign "+" ("-") corresponds to $\lambda > 0$ ($\lambda < 0$),
 the characteristic lengths $\lambda_{\pm}$ are given by
\begin{equation}
\label{lam}
\lambda_{\pm}=\sigma \Big\{\sqrt{\Big(\frac{V_{pr}(E) \; \sigma}{2 D_0}\Big)^2 + 
\frac{\sigma}{\sigma_{eff}}} \pm \frac{V_{pr}(E) \; \sigma}{2 D_0}\Big\}^{-1},
\end{equation}
where the parameter $\sigma_{eff}$ has the dimensionality of length, $\sigma_{eff}=D_0 \tau^{*}/\sigma (f + g)$, 
while the amplitudes $A_{\pm}$ in Eq.(\ref{profiles}) obey
\begin{equation}
\label{amp}
A_{\pm}=\pm \Big(\frac{V_{pr}(E) \;  \sigma_{eff}}{D_0}\Big) \frac{1}{1 + \sigma_{eff}/\lambda_{\mp}}
\end{equation}
Now several comments on the results in Eqs.(\ref{profiles}),(\ref{lam}) and (\ref{amp}) are in order. 
 Note, first, that $\lambda_{-} > \lambda_{+}$, and consequently,  the local 
density past the probe approaches its non-perturbed value $\rho_s$
slower than in front of it;  this signifies that
 correlations between the probe position and particle
distribution are stronger past the probe and demonstrates the
 memory effects of the medium. Next,  $A_{+}$ is always
 positive, while $A_{-} < 0$; this means that 
the density profile is a non-monotoneous function of $\lambda$ 
and is characterized by a jammed region in front
of the probe, in which the local density is higher than $\rho_s$, and a depleted region past the probe in which 
the density is lower than $\rho_s$.
As a matter of fact,  the depleted region appears to be more pronounced than the jammed one and
the mean density in the adsorbed monolayer perturbed by the probe 
is $\em lower$ than the mean density in the unperturbed monolayer; one readily finds that
the integral deviation $\Omega$ of the density from the equilibrium value $\rho_s$, i.e., 
\begin{equation}
\Omega= \int_{-\infty}^{\infty} \frac{d \lambda}{\sigma} \Big(k(\lambda) - \rho_s\Big)
\end{equation}
is always negative,
\begin{equation}
\label{omega}
\Omega= - \rho_s \Big(\frac{V_{pr}(E) \sigma_{eff}}{D_{0}}\Big)^2 \; 
\Big[1+\frac{\sigma_{eff}}{\sigma}+\sqrt{\frac{4 \sigma_{eff}}{\sigma} 
+\Big(\frac{V_{pr}(E) \sigma_{eff}}{D_{0}}\Big)^2}\Big]^{-1}
\end{equation}
which means that the perturbance of the monolayer created by the probe shifts the balance between
adsorption and desorption towards particles desorption into the reservoir. 
This result may be a bit counterintuitive and one could expect the 
reverse sign of $\Omega$, since 
the local density in front of the probe is higher than $\rho_s$ and particle depletion past the probe 
favors adsorption from the reservoir. However, Eq.(\ref{omega}) states that 
 the reservoir fails to equilibrate the depleted region past the probe, whose size 
$\lambda_{-}$ exceeds the size $\lambda_{+}$ of the jammed
region. Apparently, this feature may be specific to a one-dimensional system only and the sign of $\Omega$ may be different for 2D, in which
case $k(\lambda)$ shows non-monotoneous $\lambda$-dependence for $\lambda \leq 0$ in case of a lattice gas with conserved particles number 
\cite{deconinck}.
Note, lastly, that $\Omega$ in Eq.(\ref{omega}) is proportional to the second power of the probe
terminal velocity at small Peclet-type numbers 
$P = V_{pr}(E) \sigma_{eff}/D_0$, and increases linearly with $V_{pr}(E)$ when $P$ is large.

Now, we are in position to calculate the terminal velocity of the probe molecule. 
Substituting 
Eq.(\ref{profiles}) into Eq.(\ref{newvel}), we arrive at a 
closed-formed non-linear equation of the form
\begin{equation}
V_{pr}(E)=\frac{(p-q) (1-\rho_s) \sigma}{\tau} - \rho_s \sigma \frac{V_{pr}(E) \sigma_{eff}}{D_0 \tau}
\Big[\frac{p}{1+\sigma_{eff}/\lambda_-} + \frac{q}{1+\sigma_{eff}/\lambda_+}\Big]
\label{eqvelo}
\end{equation}
which determines $V_{pr}(E)$ implicitly. In Eq.(\ref{eqvelo}) the first term on the rhs is
a trivial, mean-field-type result,  which obtains in the limit of, say, perfectly stirred monolayer
with $D_0 \to \infty$ (or $\tau^{*} \to 0$). The second term on the rhs of Eq.(\ref{eqvelo}), which is proportional to the
Peclet-type number $P$,  has a more complicated origin and is associated
with collective effects - formation of a non-homogeneous, stationary density profile around the probe, 
whose characteristics  
depend themselves on the velocity.  

Eq.(\ref{eqvelo}) is rather complicated  and
can be solved for arbitrary values 
of the system parameters only numerically. 
It simplifies considerably, however, in the
limit $E \to 0$,  in which case $V_{pr}(E)$ can be
calculated analytically.  
Expanding $p$, $q$ and $V_{pr}(E)$  in
 powers of $E$ and retaining only linear in $E$
terms, 
we find that in this limit 
$V_{pr}(E)$ 
attains
the following,
physically revealing form
\begin{equation}
\label{stokes}
V_{pr}(E) \approx \zeta^{-1} E,
\end{equation}
which can be thought off as the analog of the Stokes formula for driven motion
 in a one-dimensional monolayer undergoing continuous exchanges with the vapor phase. The
friction coefficient in Eq.(\ref{stokes}) is given explicitly by
\begin{equation}
\label{friction}
\zeta = \frac{2 \tau}{\beta \sigma^2 (1- \rho_s)} \Big[1+\frac{\sigma \rho_s \sigma_{eff}}{D_0 \tau (1 +
\sqrt{\sigma_{eff}/\sigma})}\Big]
\end{equation}
Note, that $\zeta $ is a sum of two contributions. 
The first one, $\zeta_{mf}$, 
has an essentially 
 mean-field-type form and 
is just a usual
expression for the inverse mobility 
of a particle performing biased random walk on a one-dimensional lattice, 
divided by the fraction of non-occupied
lattice
sites, $(1 - \rho_s)$, which is the mean density of vacant sites 
in a monolayer with homogeneous particle distribution, i.e. monolayer 
being in equilibrium with the vapor phase. 
On contrary, the second 
term, $\zeta_{coop}$,  stems out of the cooperative effects, 
associated with formation
of a stationary, non-homogeneous density profile around the probe, and 
represents  the net resistance offered by the monolayer
particles to the probe. 
It may be instructive to single out this contribution explicitly; 
setting $\tau = 0$, 
(which means physically that the probe slides on the surface
 without dissipation regardless of the surface
corrugation),  we have 
\begin{equation}
\label{friction'}
\zeta_{coop} = \frac{2 \tau^{*} \rho_s}{\beta \sigma^2 (1- \rho_s)} 
\frac{1}{(f + g) (1 +
\sqrt{D_0 \tau^{*}/\sigma^2 (f + g)})}
\end{equation}
Comparing $\zeta_{mf}$ and $\zeta_{coop}$, we 
infer that the latter becomes 
progressively more important in the
limit, for instance, when
both  $f$ and $g$ tend to zero (while their ratio  $f/g$ is kept constant, $f/g
  = const = (1 - \rho_s)/\rho_s$).
Of course, such a behavior can be expected on physical
grounds since this limit corresponds  to a lattice gas with a fixed number of particles and suppressed 
exchanges with the vapor phase; in this case the friction coefficient (as well as 
the characteristic lengths $\lambda_{\pm}$)
is no longer constant, but rather 
diverges
as $\zeta \sim \overline{X_{pr}(t)} \sim t^{1/2}$ 
in the limit $ t \to \infty$ \cite{burlatsky,olla}.

Consider finally the situation with $E = 0$, in which case the terminal velocity vanishes 
and one expects conventional diffusive motion with the mean square displacement of the form
\begin{equation}
\label{diff}
\overline{X_{pr}^2(t)} =  2 D_{pr} t,
\end{equation}
where $D_{pr}$ is some unknown function of the system parameters. Heuristically,  we can compute  $D_{pr}$ 
for the system under study
if we assume the validity 
of the Einstein relation $\mu_{pr} = \beta D_{pr}$ between the mobility $\mu_{pr}$, 
$\mu_{pr} = lim_{E \to 0} (V_{pr}(E)/E) = \zeta^{-1}$, 
and the self-diffusion 
coefficient $D_{pr}$ of the probe particle (see for more details \cite{lebowitz2}).
We find then that the probe self-diffusion coefficient should be given  by
\begin{equation}
\label{diffusion}
D_{pr}=\frac{\sigma^2 (1- \rho_s)}{2 \tau} \Big[1+\frac{\sigma \rho_s \sigma_{eff}}{D_0 \tau (1 +
\sqrt{\sigma_{eff}/\sigma})}\Big]^{-1}
\end{equation}

In order to check our analytical predictions in Eqs.(\ref{profiles}), 
(\ref{eqvelo}) and (\ref{diffusion}), we have performed numerical
Monte Carlo simulations of the system evolution using the method of Gillespie \cite{gillespie}.
Results of these simulations, performed at different values of the parameters $f$, $g$ and $p$, 
are presented in Figs.2 to 4.    
These figures show that there is some 
systematic error 
between the analytical predictions of this Section,  Eqs.(\ref{profiles}), (\ref{eqvelo}) and  (\ref{diffusion}),
and the numerical data; this 
error seems to be small when $f$ is small but increases 
with increase of  $f$ or decrease of $g$, which means, apparently, that the error grows with the mean density of the monolayer. 
In particular, the maximal relative error $R$ appears for $g = 0.3$ and 
amounts to $5.3$, $7.7$ and $7.9$ per cent for 
Figs.3,4 and 5, respectively.
Hence, the  agreement is
quite fair, and
consequently,  Eqs.(\ref{profiles}),(\ref{eqvelo}) and (\ref{diffusion}) 
can be regarded as rather accurate
approximate results.

Now, one may attribute 
 the origin of the discrepancy
 between the numerical and analytical results
 to the following two reasons. The first is, evidently, 
the decoupling of the hierarchy of
equations for the correlation functions in Eq.(\ref{decouplage}), which allowed us to close this hierarchy at 
the level of pairwise correlations. Despite the fact that such a decoupling scheme yields  exact results
for tracer diffusion in
one-dimensional hard-core lattice gases with conserved number of particles \cite{burlatsky,olla}, it may well
be that it fails for systems in which the total number of particles is not conserved. Second, such a
discrepancy may result from the fact that here 
solution of the discrete-space equations has been found by turning to
a continuous-space limit. As we proceed to show in the next
Section, the latter point appears to be the most decisive and the solution 
of the discrete-space equations   
(\ref{vitesse}),(\ref{eqpourklambda}),(\ref{eqpourksigma}) and (\ref{eqpourk-sigma})
 is in a very good agreement with the numerical data. This  implies, in turn, that
 the decoupling in Eq.(\ref{decouplage})
is also quite a plausible assumption for the system under study.

\section{\bf Stationary solution of the discrete-space evolution equations.}

Consider now the solution of the discrete-space 
(\ref{eqpourklambda}),(\ref{eqpourksigma}) and (\ref{eqpourk-sigma}) 
in the limit $t \to \infty$. Denoting
$k_n=k(\lambda = n\sigma)$, and introducing auxiliary variables
\begin{equation}
\label{z}
z_{1}=1+\frac{p \sigma^2}{D_0 \tau} (1 - k_1)
\end{equation}
and 
\begin{equation}
\label{-z}
z_{-1}=1+\frac{q \sigma^2}{D_0 \tau} (1 - k_{-1})
\end{equation}
in order to avoid too lengthy formulae, 
we can rewrite  Eq.(\ref{eqpourklambda})
in form of the following recursion relation
\begin{equation}
\label{recursion}
z_{1} k_{n+1} - 
[z_{1}+z_{-1}+\frac{\sigma}{\sigma_{eff}}] k_{n}
+ z_{-1} k_{n-1} + \rho_s \frac{\sigma}{\sigma_{eff}} = 0,
\end{equation}
which holds for $ |n| > 1$. Equation (\ref{recursion}) has to be solved subject to the 
boundary conditions
\begin{equation}
z_{1} k_{2} - 
[z_{-1} + \frac{\sigma}{\sigma_{eff}}] k_{1} + \rho_s \frac{\sigma}{\sigma_{eff}} = 0,
\end{equation}
and 
\begin{equation}
z_{-1} k_{-2} - 
[z_{1}+\frac{\sigma}{\sigma_{eff}}] k_{-1} + \rho_s \frac{\sigma}{\sigma_{eff}} = 0
\end{equation}
Solution to these equations can be conveniently 
obtained by applying the discrete-space Fourier transformation, 
which yields an equation of essentially the same form
as Eq.(\ref{profiles}), i.e.,
\begin{equation}
\label{dprofiles}
k_{n} = \rho_s \Big[ 1 + A'_{\pm} exp\Big(-\sigma |n|/\lambda'_{\pm}\Big)\Big],
\end{equation}
with, however, different amplitudes $A_{\pm}'$ and different characteristic 
lengths $\lambda'_{\pm}$ compared to those given by Eqs.(\ref{amp}) and (\ref{lam}); 
we use here the prime to
distinguish between the discrete and the continuous-space solutions. In the discrete-space Eq.(\ref{dprofiles})
the characteristic 
lengths obey
\begin{equation}
\label{lambdas}
\lambda'_{\pm}= \; \mp \; \sigma \; ln^{-1}\Big[
\frac{z_{1} + z_{-1} + (\sigma/\sigma_{eff}) \mp \sqrt{\Big(z_{1} + z_{-1} +
 (\sigma/\sigma_{eff})\Big)^2 - 4 z_{1} z_{-1}}}{2 z_{1}}
\Big],
\end{equation} 
while the amplitudes are given respectively by
\begin{equation}
A'_{+} = \frac{z_{1} - z_{-1}}{z_{-1} - z_{1} exp(- \sigma/\lambda'_{+})} 
\end{equation}
and
\begin{equation}
A'_{-} = \frac{z_{1} - z_{-1}}{z_{-1} exp(- \sigma/\lambda'_{-}) - z_{1} } 
\end{equation}

Now, we are in position to obtain a system of equations, determining implicitly the unknown 
parameters $z_{1}$ and $z_{-1}$, which will allow us to compute
 the terminal velocity of the probe molecule; the latter is related to $z_{\pm 1}$ by 
$V_{pr}'(E) = D_0 (z_{1} - z_{-1})/\sigma$.
Substituting Eq.(\ref{dprofiles}) into Eqs.(\ref{z}) and (\ref{-z}), we find 
\begin{equation}
\label{o}
z_{1} = 1 + \frac{p \sigma^2}{D_0 \tau} \Big[1 - \rho_s - \rho_s \frac{z_{1} - z_{-1}}{z_{-1} exp(\sigma/\lambda'_{+}) - z_{1}}
\Big]
\end{equation}
and 
\begin{equation}
\label{b}
z_{-1} = 1 + \frac{q \sigma^2}{D_0 \tau} \Big[1 - \rho_s - \rho_s \frac{z_{1} - z_{-1}}{z_{-1}  - z_{1} exp(\sigma/\lambda'_{-})}
\Big]
\end{equation}
For arbitrary values of $p$, $f$ and $g$  the parameters $z_{\pm 1}$, defined by
Eqs.(\ref{o}) and (\ref{b}), and consequently, the terminal velocity $V_{pr}'(E)$ can be determined
only numerically (see Figs.3 and 4).  However, $V_{pr}'(E)$ can be found analytically in the explicit form 
in the limit of
a vanishingly  small force $E$, $E \to 0$. 
Expanding $z_{\pm 1}$ in the Taylor series in powers of $E$ and retaining
only linear with $E$ terms, i.e. 
setting 
\begin{equation}
z_{\pm 1} \approx \alpha + \alpha \gamma_{\pm 1} E,
\end{equation}
where $\alpha = 1 + \sigma^2 (1 - \rho_s)/2 D_0 \tau$,  
we find that coefficients $\gamma_{\pm 1}$ are determined by the system of two linear 
 equations:
\begin{equation}
\alpha \gamma_{1} = - \frac{\sigma^2 \rho_s}{2 D_0  \tau}  
\frac{(\gamma_{1} - \gamma_{-1})}{\Big[ exp \Big( \sigma/\lambda'_{+}(E=0)\Big) - 1\Big]} + \frac{\beta \sigma^3 (1 -
\rho_s)}{4 D_0 \tau}
\end{equation}
and
\begin{equation}
\alpha \gamma_{-1} = - \frac{\sigma^2 \rho_s}{2 D_0  \tau}  
\frac{(\gamma_{1} - \gamma_{-1})}{\Big[1 - exp\Big(\sigma/\lambda'_{-}(E=0)\Big)\Big]} - \frac{\beta \sigma^3 (1 -
\rho_s)}{4 D_0 \tau}, 
\end{equation}
which yield the Stokes-type law in Eq.(\ref{stokes}) with the friction coefficient
\begin{equation}
\label{dfriction}
\zeta' = \frac{2 \tau}{\beta \sigma^2 (1 - \rho_s)} \Big[1 + \frac{\sigma \rho_s \sigma_{eff}}{D_0 \tau}
\frac{2}{1 + \sqrt{1 + 4 \alpha \sigma_{eff}/\sigma}} \Big]
\end{equation}
Note, that similarly to Eq.(\ref{friction}) the friction coefficient $\zeta'$, calculated from the
discrete-space evolution equations, is a sum two terms; the first one 
describes a trivial, mean-field-type behavior and coincides with the result obtained within the
continuous-space approximation. The second one, which is associated with the formation of a
stationary, non-homogeneous density profile around the probe, has a different form compared to 
that in Eq.(\ref{friction}) and reduces to it
only in the limit $ \alpha \; \sigma_{eff}/\sigma \ll 1$.

Finally, Eq.(\ref{dfriction}) allows to obtain the probe self-diffusion
 coefficient explicitly. Assuming again the validity of
the Einstein relation, we find 
\begin{equation}
\label{ddiffusion}
D'_{pr} = \frac{\sigma^2 (1 - \rho_s)}{2 \tau} \Big[1 + \frac{\sigma \rho_s
\sigma_{eff}}{D_0 \tau}
\frac{2}{1 + \sqrt{1 + 4 \alpha \sigma_{eff}/\sigma}} \Big]^{-1},
\end{equation}
which has a different form  compared to Eq.(\ref{diffusion}).

Comparison between the discrete-space solution and 
numerical data  is presented in 
Figs.2 to 5. We see now that Eqs.(\ref{o}),(\ref{b}),(\ref{dprofiles})
 and (\ref{ddiffusion}) 
agree with the Monte-Carlo results essentially better than their continuous-space counterparts;  
the maximal relative error is now independent of $g$ and $f$, 
and amounts to only $1.9$, $0.8$ and $3$ per cent for Figs.3,4 and 5, respectively. We note finally that despite this good agreement 
we certainly can not
claim that Eqs.(\ref{o}),(\ref{b}),(\ref{dprofiles})
 and (\ref{ddiffusion}) provide an exact solution of the model, which can be approached apparently 
by the method developed in  Ref.\cite{olla}.

\section{Conclusions.}

To conclude,  we have studied  dynamics of a driven probe molecule in a one-dimensional
adsorbed  monolayer composed of mobile, 
hard-core particles undergoing continuous exchanges 
with the vapor phase. Within the framework of a decoupling procedure of 
 Ref.\cite{burlatsky}, based on the decomposition of the
third-order correlation functions into a product of pairwise correlations,  
we have derived dynamical discrete-space equations describing evolution 
of the density profiles, as seen from the  moving
probe, and its velocity $V_{pr}(E)$. 
These equations 
have been solved both in the continuous-space (diffusion) 
limit, as well as in their original 
discrete-space form, which allowed us to check 
the appropriance of the continuous-space description. In both cases we have determined
the probe particle terminal velocity $V_{pr}(E)$ implicitly, 
as a solution of non-linear
equations relating its value to the system parameters. We have
 shown that the discrete-space 
solution provides a very good agreement between analytical and numerical results, while
the continuous-space approach can be regarded only as a rather accurate approximation.
Further on, we have found that  
in the limit of a vanishingly small driving force
the probe velocity 
attains the form  $V_{pr}(E) \approx E/\zeta$, where $E$ is the driving force and 
the friction coefficient $\zeta$ is expressed through 
the microscopic parameters characterizing the system under study. 
This result establishes the frictional drag force exerted on the probe
by the monolayer particles in the low-$E$ limit and thus can be thought off as
the analog of the Stokes' law for the one-dimensional monolayer in contact with the vapor phase.
Lastly, we have determined explicitly the
 self-diffusion coefficient $D_{pr}$ of the probe, which result has been also  
confirmed by Monte Carlo simulations.

\vspace{0.4in}

{\Large  \bf Acknowledgments.}

The authors wish to thank G.Tarjus, J.Talbot and J.Lyklema
 for  helpful discussions. This work was supported in part by the 
French-German collaborative research
program PROCOPE.

\section{Figure Captions}

Fig.1. One-dimensional lattice partially occupied by identical hard-core particles 
(filled circles) undergoing exchanges with the reservoir - the vapor phase.
 $g$, $f$ and $l$, $l = (1 - g)/2$, denote respectively 
particles desorption, adsorption and hopping probabilities. 
The open circle denotes the probe molecule, whose hopping probabilities are $p$ and $q$, 
respectively.

Fig.2. Density profile around stationary moving probe molecule for $f = 0.1$, $g = 0.3$ and $p = 0.98$. 
The solid line is the plot of the solution, Eq.(\ref{dprofiles}),
of the discrete-space evolution equations. The empty triangles 
denote the corresponding solution in the continuous-space limit.
Filled squares are the results of Monte-Carlo simulations.

Fig.3. Terminal velocity of the probe molecule as a function of the adsorption probability $f$
at different values of the parameter $g$. The probe hopping probabilities are $p = 0.6$
and $q = 0.4$. The solid lines  give the solution of Eqs.(\ref{o}) and (\ref{b}), the dashed lines - of Eq.(\ref{eqvelo}),
while the filled squares denote the results of Monte-Carlo simulations. Upper curves correspond to $g = 0.8$, the intermediate -
to $g = 0.5$ and the lower - to $g = 0.3$, respectively.

Fig.4. Terminal velocity of the probe molecule as a function of the adsorption probability $f$. Notations
and values of $g$ are the same as in Fig.3 except that the probe hopping 
probabilities are $p = 0.98$ and $q = 0.02$.

Fig.5. Self-diffusion coefficient of the probe molecule as a function of the adsorption probability $f$. 
Notations and values of
$g$ are the same as in Figs.3 and 4.

\end{document}